\documentclass[sigconf]{acmart}

\usepackage{adjustbox}
\usepackage{amsmath,bm,bbm}
\usepackage[ruled,lined,linesnumbered, commentsnumbered, longend]{algorithm2e}
\usepackage[justification=justified,skip=3pt]{caption}
\usepackage{pgfplots}
\usepackage{multirow}
\usepackage{tikz}
\usetikzlibrary{matrix}
\usepackage{amsmath}

\makeatletter
\def\BState{\State\hskip-\ALG@thistlm}
\makeatother
\usepackage{tkz-euclide}
\usetikzlibrary{shapes.arrows, fadings, automata,tikzmark,decorations.pathreplacing,patterns}
\usetikzlibrary{intersections}
\usetikzlibrary{arrows.meta}
\usetikzlibrary{positioning, fit, mindmap, trees, calc,tikzmark,shapes}

\definecolor{quartz}{RGB}{219,223,238}
\definecolor{spring_sun}{RGB}{242,243,195}
\definecolor{dairy_cream}{RGB}{254,226,189}
\definecolor{surf_crest}{RGB}{205,230,208}
\definecolor{french_pass}{RGB}{195,232,246}
\definecolor{cosmos}{RGB}{248,209,210}
\definecolor{portafino}{RGB}{245,237,160}
\definecolor{sail}{RGB}{163,205,235}
\definecolor{hint_green}{RGB}{226,246,209}
\definecolor{bittersweet}{RGB}{255,111,105}
\definecolor{java}{RGB}{2,190,196}
\definecolor{ice_cold}{RGB}{169,232,220}
\definecolor{bgc}{RGB}{245,245,245}
\definecolor{tuatara}{RGB}{67, 67, 67}
\definecolor{aluminum}{RGB}{153,153,153}
\definecolor{silver}{RGB}{191,191,191}
\definecolor{platinum}{RGB}{228,228,228}
\definecolor{mercury}{RGB}{230,230,230}
\definecolor{gallery}{RGB}{240,240,240}
\definecolor{free_speech_aquamarine}{RGB}{0, 156, 114}
\definecolor{sun_shade}{RGB}{255, 144, 68}
\definecolor{fern}{RGB}{101,197,117}
\definecolor{french_blue}{RGB}{0, 112, 182}
\definecolor{matisse}{RGB}{25, 104, 167}
\definecolor{sushi}{RGB}{117, 168, 47}
\definecolor{shakespeare}{RGB}{85, 154, 193}
\definecolor{egg_shell}{RGB}{238, 234, 215}
\definecolor{carnation}{RGB}{245, 80, 86}
\definecolor{flamingo}{RGB}{237, 88, 85}
\definecolor{jet_stream}{RGB}{188, 214, 210}
\definecolor{jelly_bean}{RGB}{45, 126, 150}
\definecolor{tree_poppy}{RGB}{246, 154, 27}
\definecolor{deep_carmine_pink}{RGB}{236, 50, 67}
\definecolor{copper_rust}{RGB}{155, 64, 74}
\definecolor{midnight}{RGB}{0, 29, 50}
\definecolor{chilean_fire}{RGB}{215, 87, 44}
\definecolor{puerto_rico}{RGB}{94, 194, 166}
\definecolor{japanese_laurel}{RGB}{53, 116, 40}
\definecolor{fire_engine_red}{RGB}{206, 37, 51}
\definecolor{ku_crimson}{RGB}{243, 0, 25}
\definecolor{turmeric}{RGB}{211, 178, 76}
\definecolor{tahiti_gold}{RGB}{223, 102, 36}
\definecolor{outrageous_orange}{RGB}{255, 100, 45}
\definecolor{crusta}{RGB}{254, 127, 44}
\definecolor{safety_orange}{RGB}{254, 106, 0}
\definecolor{pigment_green}{RGB}{0, 175, 79}
\definecolor{jaffa}{RGB}{240, 131, 58}
\definecolor{jet_stream}{rgb}{0.69,0.61,0.85}
\definecolor{jelly_bean}{rgb}{0.47,0.32,0.66}
\definecolor{azalea}{RGB}{251, 196, 196}
\definecolor{sundown}{RGB}{249, 180, 181}
\definecolor{light_coral}{RGB}{244, 127, 123}
\definecolor{wewak}{RGB}{244, 143, 150}

\definecolor{biscay}{RGB}{44, 62, 80}
\definecolor{carmine_pink}{RGB}{231, 76, 60}
\definecolor{athens_gray}{RGB}{236, 240, 241}
\definecolor{celestial_blue}{RGB}{52, 152, 219}
\definecolor{curious_blue}{RGB}{41, 128, 185}
\definecolor{my_sin}{RGB}{255, 176, 59}
\definecolor{viridian}{RGB}{70, 137, 102}
\definecolor{tomato}{RGB}{255, 97, 56}
\definecolor{mountain_meadow}{RGB}{0, 163, 136}
\definecolor{padua}{RGB}{121, 189, 143}
\definecolor{killarney}{RGB}{56, 113, 66}
\definecolor{ocean_green}{RGB}{79, 176, 112}
\definecolor{pastel_green}{RGB}{107, 227, 135}
\definecolor{chinook}{RGB}{163, 232, 178}
\definecolor{cosmic_latte}{RGB}{222, 247, 229}
\definecolor{chateau_green}{RGB}{69, 191, 85}
\definecolor{RoyalBlue}{RGB}{69, 191, 85}

\definecolor{blue0}{RGB}{240,249,232}
\definecolor{blue1}{RGB}{204,235,197}
\definecolor{blue2}{RGB}{168,221,181}
\definecolor{blue3}{RGB}{123,204,196}
\definecolor{blue4}{RGB}{78,179,211}
\definecolor{blue5}{RGB}{43,140,190}
\definecolor{blue6}{RGB}{8,88,158}

\definecolor{yellow0}{RGB}{255,255,212}
\definecolor{yellow1}{RGB}{254,227,145}
\definecolor{yellow2}{RGB}{254,196,79}
\definecolor{yellow3}{RGB}{254,153,41}
\definecolor{yellow4}{RGB}{236,112,20}
\definecolor{yellow5}{RGB}{204,76,2}
\definecolor{yellow6}{RGB}{140,45,4}

\begin{document}

\title{End-to-End User Behavior Retrieval in Click-Through Rate Prediction Model}


\author{Qiwei Chen}
\authornote{Both authors contributed equally to this research.}
\email{chenqiwei.cqw@alibaba-inc.com}
\orcid{1234-5678-9012}
\author{Changhua Pei}
\authornotemark[1]
\email{changhua.pch@alibaba-inc.com}
\affiliation{%
  \institution{Alibaba Group.}
   \state{Beijing}
   \country{China}
}

\author{Shanshan Lv}
\email{lss271346@alibaba-inc.com}
\affiliation{%
  \institution{Alibaba Group.}
  %\streetaddress{1 Th{\o}rv{\"a}ld Circle}
  %\city{Chaoyang Qu}
  \state{Beijing}
  \country{China}}

\author{Chao Li}
\email{ruide.lc@alibaba-inc.com}
\affiliation{%
  \institution{Alibaba Group.}
  %\city{Chaoyang Qu}
  \state{Beijing}
  \country{China}}

\author{Junfeng Ge}
\email{beili.gjf@alibaba-inc.com}
\affiliation{%
 \institution{Alibaba Group.}
 %\streetaddress{Rono-Hills}
  %\city{Chaoyang Qu}
  \state{Beijing}
  \country{China}}

\author{Wenwu Ou}
\email{santong.oww@taobao.com}
\affiliation{%
  \institution{Alibaba Group.}
  %\streetaddress{30 Shuangqing Rd}
  %\city{Chaoyang Qu}
  \state{Beijing}
  \country{China}}

\renewcommand{\shortauthors}{}

\newcommand{\reed}[1]{{\color{red}{(reed: #1)}}}
\newcommand{\tbd}[1]{{\color{red}{(#1)}}}

\begin{abstract}
Click-Through Rate (CTR) prediction is one of the core tasks in recommender systems (RS). It predicts a personalized click probability for each user-item pair. Recently, researchers have found that the performance of CTR model can be improved greatly by taking user behavior sequence into consideration, especially long-term user behavior sequence. The report on an e-commerce website shows that 23\% of users have more than 1000 clicks during the past 5 months. Though there are numerous works focus on modeling sequential user behaviors, few works can handle long-term user behavior sequence due to the strict inference time constraint in real world system. Two-stage methods are proposed to push the limit for better performance. At the first stage, an auxiliary task is designed to retrieve the top-$k$ similar items from long-term user behavior sequence. At the second stage, the classical attention mechanism is conducted between the candidate item and $k$ items selected in the first stage. However, information gap happens between retrieval stage and the main CTR task. This goal divergence can greatly diminishing the performance gain of long-term user sequence. In this paper, inspired by Reformer, we propose a locality-sensitive hashing (LSH) method called ETA (End-to-end Target Attention) which can greatly reduce the training and inference cost and make the end-to-end training with long-term user behavior sequence possible. Both offline and online experiments confirm the effectiveness of our model. We deploy ETA into a large-scale real world E-commerce system and achieve extra 3.1\% improvements on GMV (Gross Merchandise Value) compared to a two-stage long user sequence CTR model.

\end{abstract}

\iffalse

\fi


\maketitle

\section{Introduction}
Recommendation systems (RS) are widely deployed to address the information overload problem. Among all those deep learning models used in RS, Click-Through Rate (CTR) prediction model is one of the most important one. Both industry and academy pay much attention on improving the AUC (area under the ROC curve) of CTR model in order to improve the online performance of RS. In the past decade, the performance of CTR model has been improved greatly. 
%with two remarkable milestones. The first milestone is the introducing of \textit{feature interaction}. These models\cite{rendle2010factorization,he2014practical,cheng2016wide,zhang2016deep,xiao2017attentional,wang2017deep,guo2017deepfm,qu2016product,lian2018xdeepfm} focus on learning interaction patterns from the existing samples by designing complex feature interaction architecture. 
One of the remarkable milestones is the introducing of \textit{user behavior sequence}, especially the long-term user behavior sequence~\cite{zhou2018deep,zhou2019deep, pi2019practice,qi2020search, qin2020user}. According to the report of \cite{ren2019lifelong}, 23\% of users in an e-commerce website have more than 1000 clicks during the past 5 months. How to effectively utilize the massive and informative user behaviors has become more and more important, which is also the goal of this paper.

Various methods are proposed to model sequential user behavior data. Early approaches, such as the sum/mean pooling methods,  RNN-based methods~\cite{cho2014properties,cho2014learning,hochreiter1997long}, CNN-based methods~\cite{lecun1989backpropagation,tang2018personalized} and self-attention-based methods~\cite{DBLP:journals/corr/VaswaniSPUJGKP17,kang2018self} encode different length of user behavior sequence into fixed dimensional hidden vector. However, they fail to capture the dynamic local interests of an user when scoring different candidate items. These methods also introduce noises by encoding all user historical behaviors. 
To overcome the drawbacks of the global pooling methods, DIN~\cite{zhou2018deep} is proposed to generate various user sequence representations according to different candidate items by \textbf{target attention} mechanism, where the target candidate item acts as query $\bm{Q}$ and each item in the sequence acts as key $\bm{K}$ and value $\bm{V}$. However, due to the expensive computation and storage resources, DIN uses the recent 50 behaviors for target attention, which ignores the resourceful information in the long user behavior sequence and is obviously sub-optimal. 

Recently, methods such as SIM~\cite{qi2020search} and UBR4CTR~\cite{qin2020user} are proposed to capture user dynamic interests from longer user behavior sequence and become the SOTA (state-of-the-art) methods. These methods act in a two-stage way. In the first stage, an auxiliary task is designed to retrieve the top-$k$ similar items from long-term user behavior sequence, such that the top-$k$ similar items are prepared in advance. In the second stage, the \textit{target attention} mechanism is conducted between target item and $k$ items selected in the first stage. However, the information used for retrieval stage is divergent or outdated with the main CTR model. For example, UBR4CTR~\cite{qin2020user} and SIM~\cite{qi2020search} use attributes such as category to select items from user behavior sequence which share the same attribute with target candidate item, which is divergent with the target of CTR model. SIM~\cite{qi2020search} also tried to building an offline inverted index 
based on the pre-trained embedding. During training and inference, the model can search the top-$k$ ``similar'' items. But most of the CTR model is in online learning paradigm and the embedding is updated continuously. Thus the pre-trained embedding in offline inverted index is outdated compared with the embedding in online CTR model. Whether the divergent target or the outdated retrieval vector, it will prevent the long-term user behavior sequence to be full utilized.  

% As a result of information gap between the retrieval part and the main CTR model can not maximum utilizing the power of long-term user behavior sequence.

% However, there are two main drawbacks for \textit{two-stage} methods. Firstly, the performance of CTR model highly depends on the accuracy of top-$k$ similar item retrieval task of first-stage, which is unfriendly to future maintenance and upgrades. Secondly, the information disagreement of two stages may lead to performance degradation. 
% For example, SIM~\cite{qi2020search} selects behavior items which share the same category with the target item. But the category is manually designed and can not be updated with the CTR model. In SIM, another retrieval method is proposed to select top-k nearest neighbor items in vector space. However, the embedding vector comes from another pre-trained model. Similarly, UBR4CTR~\cite{qin2020user} uses a separate feature selection model in the first stage. In our paper, we aim to eliminating the information gap between two stages by integrating the long-term user behavior retrieval module into the CTR model to enable end-to-end learning. 

In this paper, we propose a method called ETA, enabling end-to-end long-term user behavior retrieval to mitigate the aforementioned information gap (\textit{i.e.,} divergent target and outdated embedding) in CTR prediction task. We use SimHash to generate a fingerprint for each item in user behavior sequence. Then the hamming distance is used to help select top-$k$ items for target attention. Our method reduces the retrieval complexity from $O(L*B*d)$ multiplication to $O(L*B)$ hamming distance calculation, where $L$ is the length of behavior sequence, $B$ is the number of candidate items to be scored by CTR model at each recommendation and $d$ is the dimension of item embedding. The reduction of complexity helps us removing the offline auxiliary model and conducting real-time retrieval during training and serving procedure. This improves the ranking improvements greatly compared with SOTA models. The contributions of our paper can be summarized in three-fold.

 % The natural way to enable end-to-end user behavior retrieval is to share the embedding with the CTR model. SIM~\cite{qi2020search} can be modified to use the embedding of CTR model itself instead of pre-trained model in first-stage retrieval. However, the retrieval of k-nearest vectors is not feasible at online environment because the present k-nearest neighbor search algorithm needs $O(L*B*d)$ times multiplications between two float numbers. $L$ is the length of behavior sequence ($\geq$1000), $B$ is the batch size of target items ($\approx$1000) and $d$ is the dimension of user embedding ($\approx$128). In SIM and UBR4CTR, this problem is bypassed via offline pre-computing. K-nearest neighbor search is conducted offline and the results are stored in an inverted index for online usage. But this is not a final solution because online deep learning schema is widely used in industry and the embedding of CTR model is updated continuously. Information gap still exists between retrieval stage and training stage because building an offline inverted index is time consuming.

% In our paper, inspired by SimHash, we propose a method which can reduce the retrieval complexity from $O(L*B*d)$ multiplication to $O(L*B)$ hamming distance calculation. The retrieval time is reduced by more than $128\times$. As a result, real-time retrieval is feasible and ranking performance is improved greatly. The contributions of our paper can be summarized in three-fold. 
\begin{itemize}
    \item We propose an \underline{E}nd-to-end \underline{T}arget \underline{A}ttention method for CTR prediction task, which is called as \textbf{ETA}. To the best of our knowledge, ETA is the first work to model the long-term user behavior sequence with CTR model in an end-to-end way.  
    \item Both offline experiments and online A/B tests show that ETA achieves significant performance improvements compared with the SOTA models. We get an extra 3.1\% improvements on GMV after deploying ETA into a large-scale real world E-commerce platform when compared with a two-stage CTR model.  
    \item Comprehensive ablation studies are conducted to reveal the hands-on practical experiences for better modeling sequential user behaviors under the limitation of inference time constraint. 
    \item Our method can also be extended to other scenario to other models which need to handle extreme long sequence, \textit{e.g.,} long sequence time-series forecasting models.
    
\end{itemize}

\section{Related Work}
CTR prediction task is one of the crucial tasks in recommender systems, online advertising and information retrieval. The CTR model predicts the probability of an user clicking on a certain target item. The output probability can be used as the ranking score for the downstream ranking tasks. The accuracy of CTR model can greatly affect the online performance of online systems. For example, in our online RS, 0.1$\%$ AUC improvement of CTR model can bring millions of real-world clicks and revenue. Tremendous works focus on improving the accuracy of CTR model in different ways, which can be divided into three categories: \textbf{feature interaction}, \textbf{user behavior sequence} and \textbf{long-term user behavior sequence}. 

\textbf{Feature Interaction:} The intuition of feature interaction is to memorize the co-occurrence pattern in the feature space together with label. For example, a feature $AND(user\_installed\_app=netflix$, $impression\_app=pandora)$ can better capture the pattern for an user who clicked or did not click a certain recommended App. A series of works are published to model the feature interactions more effectively. The representative works are FM\cite{rendle2010factorization}, FFM\cite{pan2018field}, GBDT$+$LR\cite{he2014practical}, Wide\&Deep\cite{cheng2016wide}, FNN\cite{zhang2016deep}, AFM\cite{xiao2017attentional}, DeepCross\cite{wang2017deep}, DeepFM\cite{guo2017deepfm}, PNN\cite{qu2016product} and xDeepFM\cite{lian2018xdeepfm}. Various differences can be found between any two models, \textit{e.g.}, whether the deep learning technique is used, whether the embedding of weights and features are shared or whether feature engineering is needed. 
%Because our ETA model does not belong to this category, the detailed comparisons of these models are omitted for space limitation. \reed{a little weird}

\textbf{User Behavior Sequence:} The user behavior sequence is highly personalized for each user and contains spatio-temporal user interest information. Introducing the sequential user behaviors into CTR model is a remarkable milestone. Youtube\cite{covington2016deep} uses watched video sequence and search tokens in their model to capture the user interests. To better extract user interests from the user behavior sequence, various models are proposed, including CNN~\cite{tang2018personalized,yuan2019simple}, RNN~\cite{hidasi2015session,zhou2019deep}, Attention~\cite{feng2019deep,sun2019bert4rec} and Capsule Network~\cite{li2019multi}. However, the user interest vectors learned from the above models are global for a certain user. DIN~\cite{zhou2018deep} proposes an attention based method called target attention to capture the diverse local interests for a certain user facing with different target items. 

\textbf{Long-term User Behavior Sequence:} Despite the powerful ability to capture the diverse user interests, the computation of target attention is costly. The strict limit of online inference time prevents DIN-like models from using longer user sequence. MIMN~\cite{pi2019practice} can handle long-term user behavior sequence by decoupling the user interest modeling with the rest of CTR task. The user interest vector is updated offline in an asynchronous way whenever a new behavior is observed. As there is no inference time limit offline, MIMN can model any sequence lengths theoretically. However, MIMN can not learn various user interest vectors for different target items. SIM~\cite{qi2020search} and UBR4CTR~\cite{qin2020user} defeat MIMN in CTR task and become the SOTA models. Both SIM and UBR4CTR adopt two-stage architecture to model long-term user behavior sequence. At first stage, an auxiliary task is designed to retrieval the top-$k$ similar items from long-term user behavior sequence. At second stage, target attention is conducted between target item and $k$ items selected in the first stage. 

Besides the above related works on CTR prediction task, plenty of works aim to improve the efficiency and effectiveness of transformer. Reformer~\cite{kitaev2020reformer} and Informer~\cite{zhou2020informer} are the most relevant works. However, they only focus on the optimization of classical transformer and cannot be directly used on CTR prediction task because of the strict inference time constraint for large scale online recommendation systems. 
% they use the target item related information as query to retrieval $K$ top similar items. Then these $K$ items are fed into the second stage which is similar with DIN. These two models are state-of-the-art methods which have best performance so far. In the following part of this paper, we will show that by using end-to-end user behavior sequence retrieval instead of two-stage method, our ETA model outperforms SIM and UBR4CTR in both offline and online experiments.    

\section{Preliminaries}
In this section, we first give the formulation of CTR prediction task. Then we introduce how the fingerprint of a $d$-dimensional embedding vector is generated by SimHash mechanism. 
\subsection{Formulation of CTR Prediction Task}
% CTR是经典的任务之一，负责ranking的部分， youtube系统的对使用的feature x进行了分析和归类， 使用了用户行为的序列， 有一个经典的ctr的model。 以及分析一下复杂度。看要不要提到训练和预测
CTR prediction task is widely deployed in  online advertising, recommender systems and information retrieval. It aims to solve the following problem: 

\textit{Given an impression} $j$ \textit{where an item is displayed to an user}, \textit{predict the probability of user click (labeled as $y_{j}$) using the feature vector $x_{j}$.}

\begin{equation}
    p_{j} = P(y_j=1|~x_{j};\theta); \textit{j} \in \mathcal{I}.
\end{equation}

CTR task is usually modeled as a binary classification problem. For each impression $j \in \mathcal{I}$, a binary label $y_j$ is recorded according to whether the item is clicked or not. Then CTR model is trained in a supervised way to minimize the cross entropy loss, which is shown as Equation~\ref{eq:cross_entropy}, where $N$ is the number of impressions. $x_{j}$ and $y_j$ are feature vector and label of impression $j$ respectively. $\theta$ represents the trainable parameters of CTR model. For ease of presentation, we list the notations used in this paper in Table~\ref{tab:notation}. 

\begin{equation}
\begin{split}
    \mathcal{L}_{CTR}(\theta) = - \frac{1}{N} \sum_{j=1}^{N} & ~ \Big(~ y_j*log(p_j) + (1-y_j)*log(1-p_j) \Big).
\end{split}
\label{eq:cross_entropy}
\end{equation}

\begin{figure}[t]
    \centering
    \resizebox{0.5\textwidth}{!}{
    \includegraphics{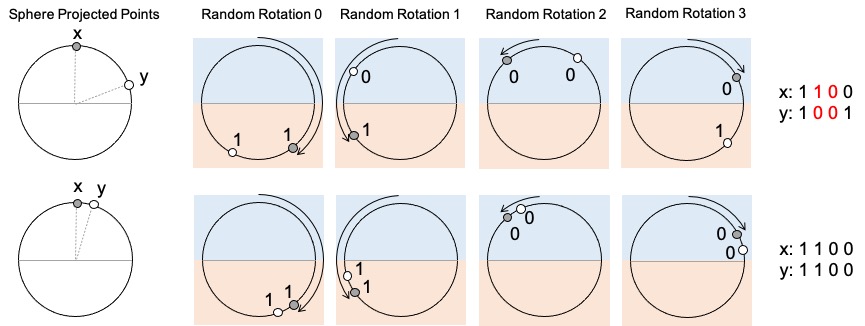}}
    \caption{Illustration of \textit{Locality-sensitive} of SimHash for two vectors $\bm{x}$ and $\bm{y}$. The detailed implementation is shown in Algorithm~\ref{algo:simhash}. Each $d$-dimensional vector is converted to a $m$-length signature vector. Each random rotation here can be regarded as one ``hash function''. The rotation is implemented by multiplying a random hash column vector $\bm{H}(i)$ (Line 2 in Algorithm~\ref{algo:simhash}). After random rotation, the spherical points are projected to signed axes (Line 3-7 in Algorithm~\ref{algo:simhash}).  Here we use 4 hash function and two projection axes to map each vector into 4 binary bits. We can observe that only those vectors who are close to each other can share more same 0/1(s), which is shown at the bottom part. }
    \label{fig:simhash}
\end{figure}

\begin{algorithm}[t]
    \SetKwFunction{isOddNumber}{isOddNumber}
    \SetKwFunction{indicator}{sgn}
    % \SetKwInput{Input}{Input}
    % \SetKwInput{Output}{Output}
    \SetKwInOut{KwIn}{Input}
    \SetKwInOut{KwOut}{Output}

    \KwIn{A $d$-dimensional embedding vector $\bm{e}_{k} \in \mathbb{R}^{1 \times d}$ \\ A fix random hash matrix $\bm{H} \in \mathbb{R}^{d \times m}$, each column \\ ~~ can be regraded as one hash function.}
    \KwOut{A binary signature vector $\bm{sig}_{k} \in \mathbb{R}^{1 \times m}$ for $\bm{e}_{k}$.}
    
    \For{$i \leftarrow 0$ \KwTo $m-1$}{
        $\bm{sig}_{k}[i] = \sum_{j=1}^{d}\indicator(\bm{e}_k[j]*H[j][i])$ \\
        \eIf{$\bm{sig}_{k}[i]>0$}{
            $\bm{sig}_{k}[i] = 1$
        }{
            $\bm{sig}_{k}[i] = 0$
        }
    }
    % $newList = [\ ]$

    % \tcc{For odd elements in the list, we add 1, and for even elements, we add 2.
    % After the loop, all elements are even.}
    % \For{$i \leftarrow 0$ \KwTo $n-1$}{
    %     \eIf{$\isOddNumber(a_i)$}{

    %         $newList.append(a_i + 1)$ \tcp*[f]{Some thought-provoking comment.}
    %      }{
    %         \tcp{Another comment}
    %         $newList.append(a_i + 2)$
    %      }
    % }

    \KwRet{ $\bm{sig}_{k}$}
    \caption{Pseudo-code of SimHash algorithm.}
    \label{algo:simhash}
\end{algorithm}

\subsection{SimHash}
\label{sec:simhash}

SimHash algorithm is first proposed by~\cite{charikar2002similarity} and one of the well-known application is~\cite{manku2007detecting} which detects duplicate web pages by SimHash based fingerprints. SimHash function takes the embedding vector of an item as input and generates its binary fingerprint. Algorithm~\ref{algo:simhash} shows the pseudo code of one possible SimHash implementation. SimHash satisfies the \textit{Locality-sensitive} properties: the outputs of SimHash are similar if the input vectors are similar to each other, which is illustrated in Figure~\ref{fig:simhash}. Each random rotation in Figure~\ref{fig:simhash} can be regarded as one ``hash function''. The rotation is implemented by multiplying the input embedding vector with a random projection column vector $\bm{H}(i)$, which is shown in Line 2 of Algorithm~\ref{algo:simhash}. After random rotation, the spherical points are projected to signed axes (Line 3-7 in Algorithm~\ref{algo:simhash}). In Figure~\ref{fig:simhash}, we use 4 hash functions and two projection axes to map each vector into a signature vector with 4 elements. Each element in the signature vector is either 1 or 0. This vector can further be decoded using an integer to save storage cost and to speed up the following hamming distance calculation. From Figure~\ref{fig:simhash}, we can observe that nearby embedding vectors can get the same hashing signature with high probability (see the bottom part of Figure~\ref{fig:simhash} compared with the upper part of Figure~\ref{fig:simhash}). This observation is the so called ``locality-sensitive'' properties. With the local sensitive properties, the similarity between the embedding vectors can be replaced by the similarity between the hashed signature. In other words, the inner product between two vectors can be replaced by hamming distance. It is noteworthy that the SimHash algorithm is not sensitive to the selection each rotation ``hashing function''. Any fixed random hash vectors are enough (see $\bm{H}(i)$ in Algorithm~\ref{fig:simhash}). It is easy to implement and can be easily applied to batches of embedding vectors.   
%For a given $d$-dimensional embedding vector $\bm{e}$, its fingerprint $\bm{h}$ is calculated using Equation~\ref{eq:simhash}. Each hash function $i$ generates a bit vector $\text{sig}_i$ as the signature of item $\bm{e}$, where each dimension is in binary manner. 
%It is noteworthy that the final bit vector $\bm{sig}_k$ for embedding vector $\bm{e}_k$  can be represented by integers.

% \begin{equation}
% \begin{split}
% \bm{h} & = \text{Int}\Big( \text{Concat}\left(\text{sig}_1,\dots, \text{sig}_i, \dots,\text{sig}_s\right) \Big). \\
% % \text{where~sig}_i &= \text{Binary}\left(\bm{w}^T\bm{H}^i\right),\\
% \end{split}
% \label{eq:simhash}
% \end{equation}

% \input{source/fig2}

\input{source/fig3}

\section{Model}

\begin{table}[t]
    \centering
    \caption{Notations used in this paper.}
    \begin{adjustbox}{max width=1.0\linewidth}
    \begin{tabular}{p{2cm}p{5cm}}
    \toprule
        Notation. & Description. \\
        \midrule
        $\mathcal{I}$ & The set of impressions.\\
        $p_j$ & The click probability of a certain impression $j$. \\
        $y_j$ & The label of click on impression $j$. \\
        $x_j$ & The feature vector on impression $j$. \\ 
        $\theta$ & The parameters of CTR model. \\
        %$j,i,u$ & The impression j where an item $i$ is displayed to user $u$ \\
        $d$ &  The dimension of item embedding. \\
        $\bm{e} \in  \mathbb{R}^{d \times 1}$ & The item embedding vector. \\ 
        $\bm{h} \in \mathbb{R}$ & The hashed fingerprint of embedding vector $\bm{e}$. \\
        $\text{sig}_i$ & The bit vector generated by $i^{th}$ hash function. \\
        $m$ & The bit-length of hashed fingerprint $\bm{h}$. \\
        %$\bm{H}$ & The seed  \\
        $\mathcal{H}_{lu}$ & Long-term user behavior sequence. \\
        $\mathcal{H}_{su}$ & Short-term user behavior sequence. \\
        $x_u$ & Raw features of user profile.  \\
        $x_c$ & Raw features of context information. \\
        $x_t$ & Raw features of target item. \\
        L & The length of long-term user behavior sequence. \\
        B & The number of candidate items to be predicted in each user request. \\
        $\bm{e}_u, \bm{e}_c, \bm{e}_t$ & The embedding vectors of user profile, context and target item respectively. \\
        $\bm{E}_s \in  \mathbb{R}^{L \times d}$ & Embedding matrix of long-term user behavior sequence.  \\
        $\bm{E}_t \in  \mathbb{R}^{1 \times d}$ & Embedding matrix of target item.  \\
        $f(\cdot)$ & The similarity function of two embedding vectors. \\
        \bottomrule
    \end{tabular}
    \end{adjustbox}
    \label{tab:notation}
\end{table}

In this section, we first introduce the detailed architecture of our \textbf{ETA} (\underline{E}nd-to-End \underline{T}arget \underline{A}ttention) model. Then we introduce different sub-modules of ETA model. At last, we introduce the hands-on experiences for the deployment of ETA.

% https://tex.stackexchange.com/questions/38868/big-parenthesis-in-an-equation
\subsection{Model Overview}
 
As shown in Figure~\ref{fig:model1}, our model takes user/item-side features as input and outputs the click probability of a certain user-item pair.
%, denoted as $p(y_i|\mathcal{H}_{lu},\mathcal{H}_{su}, x_u, x_t, x_c ; \theta)$. 
 $\mathcal{H}_{lu}$, $\mathcal{H}_{su}$, $x_u$, $x_t$ and $x_c$ are raw input features. $\theta$ represents the trainable parameters. Having these features, we use \textit{Long-term Interest Extraction Unit} (Section~\ref{sec:eu}), \textit{Multi-head Target Attention} (Section~\ref{sec:mhta}) and \textit{Embedding Layer} (Section~\ref{sec:embeddinglayer}) to convert $\mathcal{H}_{lu}$, $\mathcal{H}_{su}$, $x_u$, $x_t$ and $x_c$ into hidden vectors respectively. Then the hidden vectors are concatenated together and are fed into the MLP (Multi-layer Perception) part. At the last layer of MLP, sigmoid function is used to map the hidden vector into a scalar $p(y_j|\mathcal{H}_{lu},\mathcal{H}_{su}, x_u, x_t, x_c ; \theta)$ which represents the click probability of a certain user-item pair. This probability can be used as the ranking score for the downstream tasks. 

\subsection{Embedding Layer}
\label{sec:embeddinglayer}

For different types of features, we adopt different embedding techniques. The raw input features are mainly divided into two types: the categorical features and the numerical features. In our model, we use one-hot encoding for categorical features. For numerical features, we first divide the features into different numerical buckets. Then we apply one-hot encoding to identify different buckets, which is the similar way with~\cite{han2019all}. Note that the one-hot encoding vectors can be extremely sparse because there are billions of item ids. Thus we map all one-hot embedding vectors into low-dimensional hidden vectors to reduce the number of parameters. We use $\bm{e}_i \in  \mathbb{R}^{d \times 1}$ to represent the embedding vectors of item $i$. All the embedding vectors of user behavior items are then packed together into a matrix $\bm{E}_s \in  \mathbb{R}^{L \times d}$, which is shown in Equation~\ref{eq:embeddingmatrix}. $L$ is the length of user behavior sequence and $d$ is the embedding size. 
% In the rest part of the paper, we use $\bm{E}_s$ and $\bm{E}_t$ represent the embedding matrices for behavior sequence and target item respectively. 

\begin{align}
    \bm{E}_s &= \begin{bmatrix}
           \bm{e}_{1}^\top \\
           \bm{e}_{2}^\top \\
           \vdots \\
           \bm{e}_{L}^\top
         \end{bmatrix}.
    \label{eq:embeddingmatrix}
  \end{align}

\subsection{Multi-head Target Attention}
\label{sec:mhta}
Multi-head attention is first proposed by \cite{DBLP:journals/corr/VaswaniSPUJGKP17} and is widely applied on CTR prediction tasks~\cite{zhou2020can,xiao2017attentional,qin2020user,qi2020search,xu2020deep}. In CTR prediction task, target item acts as query ($\bm{Q}$) and each item in user behavior sequence acts as key ($\bm{K}$) and value ($\bm{V}$). We call this multi-head attention structure as \textit{multi-head target attention}, which is abbreviated as TA. 
%TA can be regarded as a special case of multi-head attention where the query sequence has only one target item. \reed{?} 
The calculation of TA is shown in Equation~\ref{eq:mhta}. The main part of TA is dot-product attention, which is shown in Equation~\ref{eq:dotattention}. The dot-product attention is made up of two steps. Firstly, similarities are calculated between each behavior item and target item according to their embedding matrices $\bm{Q}$ and $\bm{K}$. Secondly, the normalized similarities are used as the attention weights to calculate the weighted sum embedding of all the behavior items, whose embedding matrix is represented as $\bm{V}$.

\begin{equation}
\begin{split}
\text{TA}(\bm{E}_t,\bm{E}_s) & = \text{Concat}(\text{head}_1,...,\text{head}_h)\bm{W^O}, \\
\text{where~head}_i &= \text{Attention}(\bm{E}_t\bm{W}_i^Q, \bm{E}_s\bm{W}_i^K, \bm{E}_s\bm{W}_i^V),
\end{split}
\label{eq:mhta}
\end{equation}

\begin{equation}
\text{Attention}  (\bm{Q,K,V})   = \text{softmax}\left(\frac{\bm{QK}^T}{\sqrt{d_k}}\right)\bm{V},
    \label{eq:dotattention}
\end{equation}
where $\bm{E}_t \in \mathbb{R}^{1 \times d}, \bm{E}_s \in \mathbb{R}^{L \times d}$ are input embedding matrices for target item and behavior sequence respectively. $L$ is sequence length and $d$ is the embedding size of hidden vector for each behavior item. Note that we only select one piece of sample instead of a batch of samples for clarity. Matrices $\bm{Q,K,V}$ represent queries, keys and values respectively. $d_k,d_q,d_v$ is the embedding size for each row vector of $\bm{K,Q,V}$. $\sqrt{d_k}$ is used to avoid large value of the inner product. \textit{softmax} function is used to convert the value of inner-product into the adding weight of the value vector $\bm{V}$. $\bm{W}_i^Q \in \mathbb{R}^{d \times d_k},\bm{W}_i^K \in \mathbb{R}^{d \times d_k},\bm{W}_i^V \in \mathbb{R}^{d \times d_v}$. $\bm{W^O} \in \mathbb{R}^{hd_v \times d}$ is the projection matrix. $h$ is the number of headers.
% 在整个文章的某个部分，需要有一个模块说明multi-head target attention是非常花费时间和资源的

%In our online environment, $B$ equals to 1000 which is in the same scale of sequence length $L$. 

\subsection{Long-term Interest Extraction Unit}
\label{sec:eu}
This part is the main contribution of our ETA model. It extends the encoding length of user behavior sequence from tens to thousands or longer to capture the long-term user interests. As mentioned before, the complexity of multi-head target attention  is $O(L*B*d)$ where $L$ is the length of user sequence, $B$ is the number of candidate items and $d$ is the representation dimension. In large-scale online system, $B$ is close to 1000 and $d$ is close to 128. Thus directly conducting multi-head target attention on thousands of long-term user behaviors is infeasible. 
%As the inference time increase with sequence length in an linear manner
% As a result, directly conducting multi-head target attention on long-term user behavior sequence is infeasible. At online environment, we can only encode the recent $N$ ( $N \leq 150$) user behaviors into one $d$ dimension embedding. 

According to Equation~\ref{eq:dotattention}, softmax is dominated by the largest elements, for each query we only need to focus on the keys that are closest to the query, which is also confirmed by \cite{kitaev2020reformer,qi2020search,qin2020user}. Thus we can first retrieval top-$k$ items from the behavior sequence and conduct the multi-head target attention on these $k$ behaviors. However, a good retrieval algorithm should satisfy two constraints: 1) the target of retrieval part should keep the same with the whole CTR model. Only in this way can the top-k retrieved items contribute the most to the CTR model. 2) the retrieval time should satisfy the strict inference time limit to make sure the algorithm can be applied in real-world system to serve millions of request per second. We compare different retrieval algorithms in Table~\ref{tab:retrieval}. SIM~\cite{qi2020search} and UBR4CTR~\cite{qin2020user} build offline inverted index to enable 
quick searching during training and inference. However, the inputs they used for building the index are attribute information (\textit{e.g.,} category) or pre-trained embedding of items, which are different with the embedding used in CTR model. This gap violate the above constraint 1) and may lead to performance degradation. If we directly use the embedding in CTR model and search the k-nearest neighbor by inner product, $O(L*B*d)$ multiplications are needed and the inference time increases greatly. $d$ is the dimension of the embedding vector. $L$ and $B$ are the number of behavior items and target items respectively. This will violate the above constraint 2) and can not be deployed online. Our ETA uses SimHash to convert the inner product of two vectors into hamming distance calculation, which is shown in Figure~\ref{fig:model1}. This makes it possible to be deployed in real world recommendation system. Besides, the local sensitive property of SimHash ensures that the fingerprint can always keep in sync with the original embedding in CTR model. The evaluation in Section~\ref{sec:eval} shows that this compatibility can greatly improve the performance. How to choose the right hash function and the joint learning between the retrieval part and the rest part of ETA will be explained in Section~\ref{sec:selecthash} and Section~\ref{sec:joint}.  

After SimHash function and hamming distance layer, top-$k$ similar behavior items are selected from $\mathcal{H}_{lu}$, then the aforementioned multi-head target attention is conducted to generate the hidden vector. This vector acts as the representation of long-term user interest and is fed into the MLP (Multi-layer Perception) layers together with other vectors. The formulation of long-term user interest unit is shown in the following Equation~\ref{eq:longtermunit}.

\begin{table*}[]
    \centering
    \caption{Comparison between different retrieval algorithms. $d$ is the dimension of the embedding vector. $L$ and $B$ are the number of behavior items and target items respectively. $m$ is the dimension of fingerprint generated by SimHash. $M$ is the size of attribute inverted index for each user. In real world CTR model, $L=1024, B=1024, d=128, m=4, M=300$. It is not worthy that directly conduct inner product online violate the time constraint and can not be deployed in large scale online RS.}
    \begin{adjustbox}{max width=1.0\linewidth}{
    \begin{tabular}{p{4.5cm}cp{2cm}ccp{1cm}}
    \toprule
    Retrieval input.   & Retrieval method. &  Gap of goal between retrieval and CTR model.  & Retrieval complexity. & Representative. \\
    \midrule
    Attribute. & Offline inverted index & Big & $O(B*log(M))$ & SIM(hard)~\cite{qi2020search} and UBR4CTR~\cite{qin2020user}\\
    Pre-trained embedding & Offline inverted index and Euclidean distance & Medium &  $O(B*M*d)$ & SIM(soft)~\cite{qi2020search} \\
    \textbf{SimHash-based fingerprint.} & \textbf{Hamming distance} & \textbf{Small} & $O(B*L*m)$ & \textbf{ETA} \\
    Embedding of CTR model. & Inner product & No & $O(B*L*d)$ & Can not deploy.  \\
    \bottomrule
    \end{tabular}}
    \end{adjustbox}
    \label{tab:retrieval}
\end{table*}

\begin{equation}
    \text{LTI}(\bm{E}_t,\bm{E}_{s}) = \text{TA}(\bm{E}_t,\bm{E}_{s}^{'}),
    \label{eq:longtermunit}
\end{equation}

\begin{align}
   \bm{E}_{s}^{'} &= \begin{bmatrix}
           \bm{e}_{1}^\top \\
           \bm{e}_{i}^\top \\
           \vdots \\
           \bm{e}_{k}^\top
         \end{bmatrix} ,
\end{align}

\begin{align}
    \bm{e}_i & \in topk(HammingDistance(\bm{h}_i, \bm{h}_t)), \\
    \bm{h}_i & = SimHash(\bm{e}_i), \bm{h}_t = SimHash(\bm{e}_t),  
\end{align}
where LTI and TA is short for long-term user interest extraction unit and multi-head target attention respectively. $\bm{E}_{s} \in \mathbb{R}^{|\mathcal{H}_{lu}| \times d}$ is the embedding matrix of long-term user behavior sequence $\mathcal{H}_{lu}$. $\bm{E}_{s}^{'} \in \mathbb{R}^{k \times d}$ consists of top $k$ rows selected from $\bm{E}_{s_1}$ which have the largest hamming distances with target item $\bm{E}_{t} \in \mathbb{R}^{1 \times d}$.

%In our paper, three key modifications are made to enable long-term user interest extraction. (i) Filter out those behavior items which are less irrelevant with the target item, which is called as the  \textit{retrieval net}. According to Equation~\ref{eq:dotattention}, only those top similar items contribute to the final weighted sum embedding, which is also confirmed by \cite{kitaev2020reformer,qi2020search,qin2020user}. (ii) Share embedding vectors between the retrieval net and the CTR model. (iii) Use \textit{SimHash} and \textit{Hamming Distance} to enable real-time retrieval. (ii) and (iii) ensure the retrieval net always keep in sync with the CTR model, especially those online learning based CTR models.  

% TODO: 
% 1. 要不要列各种transformer的优化的文章
% 2. 要不要通过一个attention的图，来说明 在长序列中， 贡献度的分配情况。

\paragraph{Similarity Function} As shown in Figure~\ref{fig:model1}, we use SimHash function and hamming distance to calculate the similarity of two embedding vectors instead of inner product. SimHash function takes the output of aforementioned embedding layer as input. For each input embedding vector, the SimHash function generates its compressed number as the fingerprint. SimHash satisfies the \textit{locality sensitive} properties: the hashing output is similar if the input features are similar to each other. Thus, the similarity between the embedding vectors can be replaced by the similarity between the hashed fingerprints. A $d$-dimensional embedding vector can be encoded into $m$-bit number. Then the similarity between two fingerprints can be measured by hamming distance. 

%The formal definition of SimHash can be seen in Equation~\ref{eq:simhash}. 

\paragraph{Top-K Retrieval} The top-$k$ retrieval layer can find the top-k similar user behavior items with target item more efficiently by hamming distance compared with the inner-product based models. The hamming distance for two integers is defined as the number of different bits positions at which the corresponding bits are different. To get the hamming distance of two $m$-bit numbers, we first conduct $XOR$ and then count the number of bits in 1. If we define the multiplication as the atomic operation, then the complexity of hamming distance for two $m$-digit numbers is $O(1))$. 
The total complexity of the hamming distance based top-$k$ retrieval is $O(L*B*1)$, where $L$ is the sequence length, $B$ is the number of candidate items. It is noteworthy that the hashed fingerprints can be stored in an embedding table with the model each time the SimHash function is conducted. During the inference time, only embedding lookup is needed and its complexity is negligible.

\subsection{Deployment}
In this section, we show how ETA is trained with the retrieval part. Then we introduce how to select the ``hash function'' used in SimHash algorithm. Then we introduce the engineer optimization tricks.
% and compare the time complexity between different models in Table 2 %~\ref{tab:compare}.

\subsubsection{Joint learning of retrieval part}
\label{sec:joint}
During the training stage, the retrieval part does not need updates of gradients. The goal of retrieval is to select the nearest neighbors of the keys to query for the following multi-head target attention part. After the top-k closest keys to query are selected, the normal attention and back propagation are conducted on the original embedding vectors of these top-k items. The only thing for retrieval is to initialize a fix random matrix $\bm{H} \in \mathbb{R}^{d \times m}$ (see Algorithm~\ref{algo:simhash}) at the beginning of training. As long as the input embedding vector $\bm{e}_k \in \mathbb{R}^{1 \times d}$ is updated, the signature of SimHash is updated correspondingly. The \textit{Locality-sensitive} properties ensure that the top-k nearest keys to query in each iteration are selected using the latest embedding of CTR model seamlessly. Thus the gap of goal between retrieval and CTR model is much smaller than those other retrieval methods, \textit{e.g.,} offline inverted index based method shown in Table~\ref{tab:retrieval}. From the perspective of the CTR model, the retrieval part is transparent but can ensure the model use the most closet items to conduct multi-head attention. The evaluation section (Sec.~\ref{sec:eval}) shows that this end-to-end training without any pre-training or offline inverted index building can greatly improve the performance of CTR prediction task.

\subsubsection{Selection of ``Hash Function''}
\label{sec:selecthash}
The SimHash is a well-known locality-sensitive hashing (LSH)~\cite{andoni2015practical} algorithm. The implementation of SimHash is shown in Algorithm~\ref{algo:simhash} where we use fix random hash vector to act as ``hash function''. Any traditional hashing functions which hash the string to a random integer can also be used. However, in our algorithm we choose random hash vector and the implementation of Algorithm~\ref{algo:simhash} for the consideration of scalability and efficiency for matrix computation, which is the same with Reformer~\cite{kitaev2020reformer}. The locality-sensitive hashing is implemented by random rotation and projection. The random rotation refers to the multiplication between the embedding vector and a fix random hashing vector $\bm{H}(j)$. Any random $d$-dimensional vector can be used here. It is noteworthy that as we need to project the result of inner product into two singed axes to get binary signature, the element in $\bm{H}(j)$ should be generated randomly around 0.

\subsubsection{Engineered Optimization Tricks}
% 这里暂时不提memory限制的问题
When the model is deployed online, the calculation of SimHash can be reduced one step further. For a m-length signature vector $\bm{sig}_k$ for embedding vector $\bm{e}_k$ calculated by Algorithm~\ref{algo:simhash}, we can use $log(m)$-bits integer to represent the signature vector because each element in $\bm{sig}_k$ is either 1 or 0. This can greatly reduce the cost of memory and can speed up the calculation of hamming distance. The calculation time of two integers can be conducted in $O(1)$ time complexity and can be neglected. 

\begin{table}[t]
    \centering
    \caption{Corpus sizes of dataset used in this paper.}
    \begin{adjustbox}{max width=1.0\linewidth}
    \begin{tabular}{ccccc}
    \toprule
         & Users & Items &  Categories & Instances \\
        \midrule
        %Amazon(Book) & 75,053 & 358,367 & 1,583 & 22,507,155 \\
        %\midrule
        Taobao. & 987,994 & 4,162,024 &  9,439 & 100,150,807 \\
        %\midrule
        Industrial(Our Own) & 0.4 billion & 0.7 billion & 24,568 & 142 billion \\
        \bottomrule
    \end{tabular}
    \end{adjustbox}
    \label{tab: ranking}
\end{table}

% 放一个多阶段的演进的图， 最左边经典youtube， 然后是dnn， 然后是我们的模型图

\section{Experiments}
\label{sec:eval}
In this section, we conduct experiments to answer the following research questions.

\begin{itemize}
% target attention的重要性
    %\item \textbf{RQ1}: Is multi-head target attention necessary in CTR prediction tasks? 
% 长序列做target attention的重要性， 说明为什么要初筛备选item的重要性
% 在初筛item，然后做top50 target attention的方法中，各个的性能差距。为公平比较，选择的candidate set应该是一样大小的。  数据集合保持一致就可以。
    \item \textbf{RQ1}: Does our ETA model outperform the baseline models?
    %on two public datasets, one industrial dataset and online A/B test? 
% cost的大小，长序列重要的一点
    \item \textbf{RQ2}: What is the inference time of our ETA model compared with the baseline models? Inference time is as important as performance because it decides whether the model can be deployed online for serving. 
% 在不同的candidate set大小下，性能的变化。 inference长度的变化
    \item \textbf{RQ3}: Which part of our ETA model contributes the most to the performance and inference time? 
    %\item \textbf{RQ3}: How does the performance and inference time of ETA model vary with different retrieval size and the number of hash functions? 
\end{itemize}
Before presenting the evaluation results, we first describe the datasets, baseline models, metrics and experimental settings. 

\subsection{Datasets}
To conduct comprehensive comparisons between our ETA model and the baseline models, both public dataset and industrial dataset are used. Online A/B test is also conducted. For public dataset, we choose Taobao dataset, which is also adopted by baseline models SIM~\cite{qi2020search} and UBR4CTR~\cite{qin2020user}. An industrial dataset is prepared as the supplement for public dataset. Table~\ref{tab: ranking} gives a brief introduction of the datasets. 

%The detailed pre-processing of the datasets is introduced in Section~\ref{sec:setting}.

% \textbf{Amazon Book Dataset~\footnote{http://jmcauley.ucsd.edu/data/amazon/}}: This dataset is made up of users' reviews for the books in Amazon website. The dataset contains over 22 million instances. In average, each user has about 300 reviews and each book receive about 63 reviews. The review action is regarded as user behavior. For each user, the average length of user behavior sequence is 2.8. 

%TODO 用的features也要说一下
% Table~\ref{tab: ranking} lists the meta features of user and item we used in this paper. 
%TODO 序列长度要说
% We use the same pre-processing method with one of our baseline SIM\cite{qi2020search} for better comparison. (i). Sorting the reviews by the timestamp for each user. (ii). Split the sequence, take the recent 16 reviews as short-term user behavior sequence and take the rest as the long-term user sequence.  
% The category feature makes it possible to reproduce the results of SIM-hard algorithm. 

\textbf{Taobao Dataset~\footnote{\url{https://tianchi.aliyun.com/dataset/dataDetail?dataId=649&userId=1}}}: This dataset is first released by~\cite{zhu2019joint} and is widely used as the public benchmark for CTR prediction task and sequential recommendation task. It is made up of users' behavior logs from Taobao Mobile App. The user behaviors include click, favorite, add to cart and buy. This dataset contains 100 million instances. In average, each user has about 101 interactions and each item receives over 24 interactions. The recent 16 behaviors are selected as short-term user behavior sequence and the recent 256 behaviors are selected as long-term user behavior sequence. 
% There are 9439 categories in total. For each category, the maximum, minimum, mean and median number of reviews are 4,862,531, 1, 10,610 and 304 respectively.

\textbf{Industrial Dataset~\footnote{This dataset will be released to public to help the research on long-term user interest modeling.}}: This dataset is collected from our own online RS, which is one of the top-tier mobile Apps in our country. There are three advantages for our industrial dataset. (i) Our dataset contains impression interaction, which indicates an item is displayed to an user but not clicked by the user. Impression interaction is naturally the negative sample of CTR model. As a result, the tricky negative sampling is not needed. (ii) The user behavior sequence is much longer in our industrial dataset. There are over 142 billion instances and the average length reaches 938, which is 9 times longer than the public Taobao dataset. (iii) Our industrial dataset has more features designed by multiple software engineers which is closer to the real-world RS models. The recent 48 behaviors are selected as short-term user behavior sequence and the recent 1024 behaviors are selected as long-term user behavior sequence. In the ablation study, we also try the long-term user behavior sequence with the lengths in $\{256, 512, 2048\}$.

\subsection{Baselines and Metrics}
\label{sec:baselines}
\textbf{Baselines:}
We compare our model with the following mainstream baselines for CTR prediction. Each baseline is chosen to answer one or more related research questions mentioned above.  

\begin{itemize}
    
    \item \textbf{Avg-Pooling DNN}: The simplest way to utilize user behavior sequence is average pooling which encodes the various length of user sequences into fixed-size hidden vector. This baseline can be regarded as variant of DIN by replacing the target attention with average pooling, which is similar to YouTube~\cite{covington2016deep}. This baseline is mainly used to show the necessity of target attention when compared with DIN.  
    \item \textbf{DIN}~\cite{zhou2018deep}: DIN is proposed to model personalized user interests with different target items by an attention mechanism, which is called as target attention. However, DIN only utilizes the short-term user behavior sequence.
    \item \textbf{DIN (Long Sequence)} is DIN equipped with long-term user behavior sequence $\mathcal{H}_{lu}$. $\mathcal{H}_{lu}$ is encoded by mean pooling. This baseline is used to measure the information gain of long-term user behavior sequence itself when compared with DIN. 
    \item \textbf{SIM(hard)}~\cite{qi2020search}: SIM is the CTR prediction model which propose a search unit to extract user interest from long-term user behavior sequence in a two stage manner. SIM(hard) is SIM which searches the top-k behavior items by category id in the first stage.    
    \item \textbf{UBR4CTR}~\cite{qin2020user}: UBR4CTR is also a two-stage method which utilizes the long-term user behavior sequence in CTR prediction task. In UBR4CTR, a query is prepared by a feature selection model to retrieve the most similar behavior items. An inverted index is prepared for online usage. As UBR4CTR and SIM are published almost simultaneously, they do not compare to each other. In our paper, we compare both UBR4CTR and SIM for the first time. 
    \item \textbf{SIM(hard)/UBR4CTR + timeinfo} is SIM(hard)/UBR4CTR with time embedding when encoding the user behavior sequence.
\end{itemize}
% \vspace{-50pt}

In~\cite{qi2020search}, the authors propose SIM(soft) as the variants of base algorithm SIM(hard). They finally adopt SIM(hard) method as their online serving algorithm and deploy SIM(hard)+timeinfo online to serve the main traffic. This is because SIM(hard) does not need pre-training and is more friendly to system evolution and maintenance. Besides, SIM(hard) + timeinfo can achieve comparable performance with SIM(soft). Thus, we choose SIM(hard) and SIM(hard) + timeinfo as our strong baselines.

% 这儿可以有个比较的表格
MIMN~\cite{pi2019practice} is proposed by the same team with DIN. A multi-track offline user interest center is proposed by MIMN to extract the long-term user interest. At the publishing time, it achieves the state-of-the-art performance by leverage the long-term user behavior sequence. However, MIMN is defeated by SIM~\cite{qi2020search} from the same team. As MIMN contributes little to our research questions, we omit this baseline for the space limitation. 

\textbf{Metrics:} For offline experiments, we adopt widely used \textit{area under ROC curve} (AUC) as our main metric and \textit{Inference Time} as a supplement metric. AUC is suitable for binary classification problem for measuring pairwise ranking performance. \textit{Inference Time} is defined as the round-trip time when scoring a batch of items for a certain model request. We measure the \textit{Inference Time} by deploying models online to serve user requests which are copied from product environment. The machines and the number of user requests are controlled same for fairness comparison.  

 For online A/B test, we use CLICK and CTR as evaluation metrics. CLICK is defined as the total number of clicked items. CTR is used to measure the willingness of click for users in the platform. It is defined as $CLICK/PV$ where PV is defined as the total number of displayed items. 
 
% \begin{equation}
%     \text{CTR} = \frac{\text{CLICK}}{\text{PV}} = \frac{\sum_{j \in \mathcal{I}}{\mathbbm{1}(j)}}{|\mathcal{I}|}
% \label{eq:ctr}
% \end{equation}

% where $\mathcal{I}$ is the set of all impressions. $\mathbbm{1}(j)$ is the indicator function whether item is clicked in a certain impression $j$. CLICK (PV) is defined as the total number of clicked (displayed) items. 

% IPV is a supplement metric for CTR. A strategy is proved to be better if and only if improvements are observed at both CTR and IPV metrics.

% 训练集和validation和test的分割要放在这里说明
\subsection{Experimental Setup}
\label{sec:setting}
In this section, we first introduce the pre-processing for offline datasets. Then we list the hyper-parameters of both baselines and our models.

Taobao Dataset only contains positive interactions, \textit{e.g.}, review, click, favorite, add to cart and buy. We use the same data pre-processing method with MIMN~\cite{pi2019practice}. Firstly, for each user, we select the last behavior as the positive sample. Then we randomly sample a new item with same category as the negative sample for this user. 
%Then a random number $r \in [0,1]$ is generated. If $r$ is smaller than a certain threshold, we assign this candidate sample a positive label, else we 
The rest behavior items are used as features. The samples are split into training set (80\%), validation set (10\%) and test set (10\%) according to the timestamp $t$ of this sample.

Our Industrial dataset has positive and negative samples naturally because we log all the impressions for each user. An impression is labeled as positive if the item is clicked by the user. Otherwise, it is labeled as negative sample. We use the past two weeks' logs as training set and the following day as the test set, which is similar with SIM~\cite{qi2020search}. 
% It is noteworthy that we down sample the negative samples used by Facebook~\cite{he2014practical} for speeding up the training procedure without sacrifice the accuracy. 
% The online A/B test environment is carefully designed for years by our engineering team. Each bucket contains millions of users and any improvement larger than 1$\%$ on CTR or IPV is regarded as an significant result. 

% 这里酌情看要不要写的那么详细， 先列个表，后面再说
For each model on different datasets, we use the validation set to tune the hyper-parameters to get the best performance. The learning rate is searched from $1\times10^{-4}$ to $1\times10^{-2}$. The L2 regularization term is searched from $1\times10^{-4}$ to $1$. All the models use Adam optimizer. The batch size is 256, 1024 for Taobao and our industrial dataset respectively.

\begin{table}[]
    \centering
    \caption{Experimental results on Taobao Dataset.}
    \begin{adjustbox}{max width=1.0\linewidth}
    \begin{tabular}{p{4cm}c}
    \toprule
     Method   &  AUC   \\
    \midrule
        Avg-Pooling DNN & 0.8442 \\
        %\midrule
        DIN  & 0.8626  \\
        %\midrule
        DIN (Long Sequence)  & 0.8661  \\
        %\midrule
        %\midrule
        UBR4CTR &  0.8651  \\
        % UBR4CTR  & 0.8496  \\
        %\midrule
        UBR4CTR+timeinfo  &  0.8683 \\
        %UBR4CTR+timeinfo  & - \\
        %\midrule
        SIM(hard) & 0.8675 \\
        % SIM(hard) &  0.9267 \\
        %\midrule
        SIM(hard)+timeinfo &   0.8708  \\
        %SIM(hard)+timeinfo  & 0.9381 \\
        %\midrule
        ETA  & 0.8721 \\
        ETA+timeinfo   &  0.8746 \\
        \bottomrule
    \end{tabular}
    \end{adjustbox}
    \label{tab:offline}
\end{table}

\begin{table}[]
    \centering
    \caption{Experimental results on Industrial Dataset.}
    \begin{adjustbox}{max width=2.0\linewidth}
    \begin{tabular}{lcc}
    \toprule
     Method   &  AUC & Inference Time(ms)  \\
    \midrule
      Avg-Pooling DNN & 0.7216 & 8 \\
        %\midrule
      DIN & 0.7279 & 11 \\
        %\midrule
      DIN (Long Sequence)  & 0.7311 & 14 \\
        %\midrule
        %\midrule
      UBR4CTR &  0.7318 & 41 \\
      %UBR4CTR  & 0.7182 & 21 \\
        %\midrule
      UBR4CTR+timeinfo & 0.7331 & 41 \\
      %UBR4CTR+timeinfo & 0.7195 & 21 \\
        %\midrule
      SIM(hard) &  0.7327 & 21 \\
      %SIM(hard)  & 0.7345 & 21 \\
        %\midrule
      SIM(hard)+timeinfo  & 0.7338 & 21 \\
      %SIM(hard)+timeinfo & 0.7356 & 21 \\
        %\midrule
      ETA  & 0.7361 & 19 \\
      ETA+timeinfo  & 0.7373 & 19 \\
        \bottomrule
    \end{tabular}
    \end{adjustbox}
    \label{tab:our}
\end{table}

\setlength{\tabcolsep}{10pt}
\begin{table}[t]
    \centering
    \caption{Relative performance improvements and Inference Time in online A/B test compared with a DIN-based method without long-term user behavior sequence. Note that 1\% improvement on GMV is a significant improvement because it means that millions of more revenues are brought to the recommender system.}
    \begin{adjustbox}{min width=.4\linewidth}
    \begin{tabular}{cccc}
    \toprule
       Method  & CTR & GMV & Inf. Time \\
        \midrule
     %  DIN & +0 & 0 \\
        %\midrule
        SIM(hard)+timeinfo & 4.53$\%$ & 6.6$\%$ & 21ms \\
        %\midrule
        ETA+timeinfo& 6.33$\%$ & 9.7$\%$ & 19ms   \\
        \bottomrule
    \end{tabular}
    \end{adjustbox}
    \label{tab:online}
\end{table}

\subsection{Performance Comparison}
\textbf{Taobao Dataset:} The evaluation results on Taobao dataset are shown in Table~\ref{tab:offline}. From the table, we find that our ETA has stable performance improvements compared with all baselines. ETA outperforms SIM(hard) by 0.46\% and outperforms DIN(Long Sequence) by 0.6\%. 
After adding time embedding, ETA+timeinfo outperforms SIM(hard)+timeinfo by 0.38\% and outperforms DIN(Long Sequence) by 0.85\%. Similar results can be observed on SIM(hard) and UBR4CTR. 
It is observed that DIN(Long Sequence) brings 0.35\% improvement on AUC compared to DIN, which shows the effectiveness of modeling the long-term user behavior sequence for CTR prediction. 
We also find that UBR4CTR performs worse than DIN(Long Sequence). It is because that the feature selection model of UBR4CTR only selects the behaviors whose features(eg. category, weekday) are same with the target item. This kind of filtering in UBR4CTR helps removing noises away from the sequence, but can also lead to shorter user sequence, which is harmful when there are not enough items for top-k retrieval. From Table~\ref{tab:offline}, we find that DIN outperforms Avg-Pooling DNN by 1.84\%, demonstrating the fact that using target attention to encode the user sequence can greatly improve the performance. 
%In Taobao Dataset, each user has 101 behaviors in average and there are over 9,000 categories in total. As a result, the average sequence length after the first stage of UBR4CTR is 4.8, which is much shorter than 86.4 of DIN(Long Sequence). 

%in our online RS, 0.1%AUC improvement of CTR model can bringmillions of real-world clicks and revenue. 
\textbf{Industrial Dataset:} The evaluation results on our own Industrial Dataset are shown in Table~\ref{tab:our}. Note that 0.1\% AUC improvement of CTR model can bring millions of real-world clicks and revenue in our online RS. Our ETA achieves the best performance compared with all baselines. Our base ETA achieves 0.34\% and 0.43\% improvements compared with SIM(hard) and UBR4CTR respectively. Our ETA+timeinfo achieves 0.35\% and 0.42\% improvements compared with SIM(hard)+timeinfo and UBR4CTR+timinfo respectively. Different from the experimental results in public dataset, SIM(hard)+timeinfo becomes the strongest baseline on Industrial Dataset and outperforms DIN(Long Sequence) by 0.27\%. This is because of two reasons. On one hand, the user sequence length in Industrial Dataset is large enough, which is friendly to long-term user sequence based models. The average length of Industrial Dataset reaches 938, which is 9 times longer than the public Taobao dataset. On the other hand, DNN (Long Sequence) uses average pooling to encode the whole sequence without selection, which may import noises compared with retrieval based models such as UBR4CTR, SIM and ETA. 

%We can also observe the same facts with public dataset. (i) Adding timeinfo into the model can achieve better performance. For example, Sim(hard)+timestamp outperforms SIM(hard) by 0.15\% on AUC. (ii) Target attention can achieve better performance than average pooling. For example, DIN achieves 0.87\% improvements compared with Avg-Pooling DNN method. (iii) Modeling long-term user interests can improve the performance. For example, DIN(Long Sequence) achieves 0.44\% improvement compared with DIN by simply encoding the long-term user behavior sequence with average pooling method.

We can find another fact that SIM(hard) has 0.09\% performance improvement compared with UBR4CTR. This is mainly caused by different processing methods on user behavior sequence. In SIM(hard), the user sequence is split into two separate sub-sequences, which is similar with our ETA in Figure~\ref{fig:model1}. The short-term user behavior sequence $\mathcal{H}_{su}$ is made up of recent $k$ user behaviors from item 1 to item $k$. The long-term user behavior sequence $\mathcal{H}_{lu}$ is made up of another $k$ behaviors selected from item $k+1$ to item $n$. However, UBR4CTR selects a $2*k$-length behavior sequence from item $1$ to item $n$. As a result, the most recent $k$ items ($\bm{e}_1$ to $\bm{e}_k$ in Figure~\ref{fig:model1}) are selected by SIM(hard) in 100\% probability and are selected by UBR4CTR in $p$ probability decided by the feature selection model. However, timeinfo plays an important role in user interest modeling because user interest is dynamic and changes frequently. Thus SIM(hard) performs betther than UBR4CTR.
%Thus we suggest to add timeinfo into the feature selection model of UBR4CTR to improve the performance one step further. 

% The more recent the item is, the more information it contains. This is also confirmed by our ``+timeinfo'' experiment. We can observe an average 1.2\% improvements when comparing ETA/SIM(hard)/UBR4CTR with ETA/SIM(hard)/UBR4CTR+timeinfo methods respectively. In the ablation study part~\ref{sec:ablation}, we also show the performance when we merge $\mathcal{H}_{lu}$ and $\mathcal{H}_{su}$ to conduct multi-head target attention in the same way with UBR4CTR. 
 
\textbf{Online A/B Test}: The evaluation results of online A/B test is shown in Table~\ref{tab:online}. Table~\ref{tab:online} shows the performance improvements over a DIN-based method, where the DIN-based method does not have long-term user behavior sequence. From Table~\ref{tab:online}, we find that our ETA+timestamp achieves 6.33\% improvements on CTR and brings 9.7\% extra GMV compared with the DIN-based method. Compared with the strongest baseline SIM(hard)+timeinfo, our ETA+timeinfo has extra 1.8\% improvements on CTR and 3.1\% improvements on GMV. Note that 1\% improvement on GMV is a significant improvement because it means that millions of more revenues are brought to the recommender system.

% We also compare the performance with the strongest baseline SIM(hard)+timeinfo. Our ETA+timeinfo outperforms SIM(hard)+timestamp by 1.40\% on CTR and 1.7\% on CLICK with a comparable inference time (19ms v.s. 21ms). This result proves that our ETA can encode long-term user behavior sequence more effectively and efficiently comparing with SOTA model.

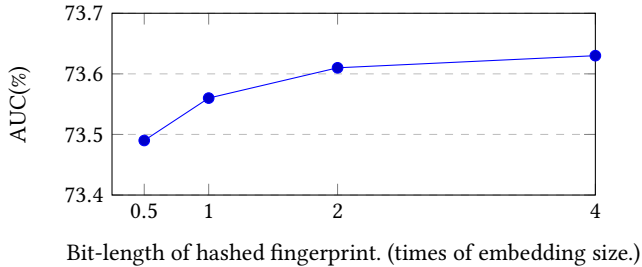
\begin{figure}
    \centering
 %   \resizebox{5cm}{2cm}{
    
\begin{tikzpicture}
\begin{axis}[
    title={},
    xlabel={Bit-length of hashed fingerprint. (times of embedding size.)},
    ylabel={AUC(\%)},
    xmin=0.25, xmax=4.0,
    ymin=73.4, ymax=73.7,
    xtick={0.5,1.0,2.0,4.0},
    ytick={73.4,73.5,73.6,73.7},
    legend pos=north west,
    ymajorgrids=true,
    grid style=dashed,
    height=4cm,
    width=0.45\textwidth
]

\addplot
    coordinates {
    (0.5,73.49)(1.0,73.56)(2.0,73.61)(4.0,73.63)
    };
%\legend{SIM,ETA-1024, ETA-2048}]
\end{axis}

\end{tikzpicture}
%  }
    \caption{AUC of ETA with different bit-lengths of hashed finger. The other model settings are the same with ETA in Table~\ref{tab:our}.}
    \label{fig:hashfunction}
\end{figure}

\setlength{\tabcolsep}{5pt}

\begin{table}[]
    \centering
    \caption{Ablation study of our ETA model on Industrial Dataset. \textbf{v0} is the base version of ETA. avg($\cdot$) and ta($\cdot$) represent encoding user behaviors by average pooling and target attention respectively. ta(1024 -s- 48) represents conducting target attention on top-48 user behaviors selected from 1024 sequential user behavior items. Symbol \textbf{s} in ta(1024 -\textbf{s}- 48) represents SimHash is used to select top-48 from 1024. Similarly, symbol \textbf{i} in ta(1024 -\textbf{i}- 48) represents inner product is used to select top-48 from 1024. }
    %\begin{adjustbox}{max width=1.0\linewidth}
    \begin{tabular}{cccc}
    \toprule
     %\shortstack[c]{ETA \\ Version}   & \shortstack[c]{Encoding \\ Manner} & \multirow{2}{*}{AUC} & \shortstack{Inference \\ Time} \\
     ETA & Encoding & \multirow{2}{*}{AUC} & Inference \\
     Version & Manner & & Time(ms) \\
    \midrule
        v0  & ta(1024 -s- 48)  & 0.7361 & 19 \\
        \midrule
        v1  & avg(1024)  & 0.7311 & 14 \\
        \midrule
        v2.1  & ta(256 -s- 48)  & 0.7339 & 14 \\
        v2.2  & ta(512 -s- 48)  & 0.7348 & 16 \\
        v2.3  & ta(2048 -s- 48)  & 0.7394 & 23 \\
        %   & ta($\infty$ -s- 48) + ta(48) & - & - \\
        \midrule
        v3  & ta(1024 -i- 48) & 0.7368 & 32 \\
       
        \midrule
        v4  & ta(1024)  & 0.7371 & 35 \\
        %\midrule
        %v5 (+timeinfo)  & ta(1024 -s- 48)  & 0.7373 & 19 \\
        \bottomrule
    \end{tabular}
    %\end{adjustbox}
    \label{tab:ablationstudy}
\end{table}

\subsection{Inference Time Comparison}
\label{sec:inftime}
Though the performance of CTR prediction is improved using long-term user behavior sequence, the model complexity increases accordingly. We measure the inference time of different models which are shown in Table~\ref{tab:offline}. Avg-Pooling DNN has the smallest inference time of 8 milliseconds (ms). It  only encodes the recent behavior items with the average pooling method. After replacing average pooling to target attention, the inference time increases by 3ms (8ms to 11ms). After importing the long-term user behavior sequence, the inference time increases by another 3ms (11ms to 14ms) . SIM and our ETA have comparable inference time around 19$\sim$21 ms. 
% However, the inference time of SIM(hard) here does not contain the offline retrieval and index building time. 
% UBR4CTR has the largest inference time because an extra IDF and BM25 based procedure is conducted online to get top-$k$ items after the retrieval stage.  
UBR4CTR has the largest inference time because an extra feature selection model is used before the retrieval stage, and an relative time-consuming IDF and BM25 based procedure is conducted online to get top-$k$ items.
%It is noteworthy that both UBR4CTR and SIM use offline inverted index to help reducing the inference time in Table~\ref{tab:offline}. If they use online retrieval instead of offline index, the inference time will be doubled to about 32ms (see Section~\ref{sec:ablation}) and it is not acceptable in online environment. Our ETA uses end-to-end online retrieval but can reach a comparable inference time compared with those offline inverted index based methods, \textit{i.e.,} UBR4CTR and SIM.  

\subsection{Ablation Study}
\label{sec:ablation}
The results of ablation study is shown in Table~\ref{tab:ablationstudy} to answer the research question RQ3. We use encoding manner to distinguish different versions (v0 to v4) of our ETA model. Note that \textbf{v0} is the base version of ETA. The encoding manners are listed in the second column of Table~\ref{tab:ablationstudy}, where avg($\cdot$) and ta($\cdot$) represent encoding user behaviors by average pooling and target attention respectively. ta(1024 -s- 48) represents conducting target attention on top-48 user behaviors selected from 1024 sequential user behavior items. Symbol \textbf{s} in ta(1024 -\textbf{s}- 48) represents SimHash is used to select top-48 from 1024. Similarly, symbol \textbf{i} in ta(1024 -\textbf{i}- 48) represents inner product is used to select top-48 from 1024. 

From Table~\ref{tab:ablationstudy}, the following observations are made. (i) Directly conduct multi-head target attention on the original 1024-length user sequence (v4) can achieve the best performance but have highest inference time at the same time. Comparing with v4, our base ETA (v0) selecting the top-$k$ behaviors for attention sacrifices about 0.1\% AUC and reduces the inference time by 46\%. (ii)  Comparing v3 with v0, replacing the SimHash with inner product at the retrieval stage achieves 0.07\% improvement on AUC. However, the inference time increases by 68\%, which is not acceptable with our strict online SLA (service level agreement). (iii) Trade-offs between AUC and inference time are observed when we change the length of user behavior sequence (v2.$x$ versus v0). The appropriate sequence length can be decided according to the requirements of online inference time. We also evaluate the performance under different bit-lengths of hashed fingerprints $\bm{h}$ generated by SimHash in Figure~\ref{fig:hashfunction}. As mentioned in Section~\ref{sec:simhash}, the bit-length of fingerprint can be controlled by the number of hash functions used in SimHash. We can find the fact that AUC can be improved by increasing the bit-length of $\bm{h}$. However, when bit-length of $\bm{h}$ is larger than $2\times$ embedding size, the improvement on AUC becomes marginal. 

% We suggest that the number of bits for hashed fingerprint should be controlled to twice large with the embedding size of CTR model because the stationary point is observed in Figure~\ref{fig:hashfunction}.

% (iiii) It is better to choose target attention on shorter user sequence instead of average pooling on longer user sequence. Comparing v2.3 with v1, conducting retrieval and target attention on 256-length sequence is better than simply averaging on 1024-length sequence can achieve 0.4\% improvements on AUC with the same inference time. Besides, from Figure~\ref{fig:hashfunction} 

%\section{Conclusion \& Future Work}

\section{Conclusion}

In this paper, we propose ETA model for CTR prediction task. To the best of our knowledge, ETA is the first method which can model the CTR together with long-term user behavior sequence in an end-to-end way. Compared to the SOTA two-stage models, the end-to-end paradigm enables the retrieval part share the information with the main part of CTR model seamlessly, which improves the prediction performance significantly. Besides, it is friendly for the maintenance and evolution of CTR model in large-scale online RS.  
To achieve the goal of end-to-end online retrieval, we propose a SimHash based method to reduce the complexity of traditional top-$k$ retrieval from $O(L*B*d)$ multiplication to $O(L*B)$ hamming distance calculation, where $L$ is the length of user sequence. $B$ is the number of candidate items for each user request. $d$ is the dimension of item embedding.
% AUC on offline dataset is improved by 0.38\% and 
Both offline and online experiments confirm the effectiveness of our ETA. The total GMV are improved by 3.1\% in online A/B test compared with the SOTA model. ETA has been deployed online to serve the mainstream traffic.

%To overcome the resource constraint when realizing the end-to-end paradigm, we adopt the SimHash method and use hamming distance instead of inner product between any item pairs. In this way, we reduce the computation complexity by xx$\times$. Both offline and online experiments are conducted to evaluate the performance of ETA model: The AUC of ETA is xxx$\%$, yyy$\%$ and zzz$\%$ higher than the SOTA model on two public dataset and our own industrial dataset. The online A/B test shows that our ETA outperforms the SOTA model by aaa$\%$ in CTR and bbb$\%$ in IPV. ETA has been deployed online to serve the mainstream traffic for our large-scale RS.
% 想要篇幅可以加回来
% In the future work, we will focus on how to reduce the calculation of hamming distance based retrieval one step further. Target attention uses target item to extract dynamic user representation with different target items. Similarly, the context information can also be used to extract dynamic user interests under different environment, \textit{i.e.,} different hours of a day, whether it is weekend or not. 

% We observe that the target attention happens between item pairs and it is possible for caching the popular middle results to speed up the model. Besides, inspired by the target attention mechanism which shared by all users and contexts, we believe that personalized target attention is also a promising direction. 

\balance
\bibliographystyle{ACM-Reference-Format}
\bibliography{sample-base.bbl}


\end{document}